\documentclass[12pt]{article}
\usepackage{epsfig}
\usepackage{amsfonts}
\usepackage{amsopn}
\usepackage{amsmath}

\title{Uniqueness of the coordinate independent $Spin(9) \times SU(2)$ state of Matrix Theory}
\author{
Mariusz Hynek\thanks{e-mail: mariusz.hynek@uj.edu.pl} \  and Maciej Trzetrzelewski\thanks{e-mail: 33lewski@th.if.uj.edu.pl}  \\ \\
 Institute of Physics,\\
Jagiellonian University, \\
Reymonta 4, 30-059 Krak\'ow,\\
Poland
}

\begin{document}
\date{}
\maketitle

\abstract{We explicitly prove, using some nontrivial identities
involving gamma matrices, that there can be only one $Spin(9)\times
SU(2)$ invariant state which depends only on fermionic variables.}

\section{Introduction}

The explicit construction of the conjectured ground state of the
Supermembrane/ M-theory matrix model \cite{model} would certainly
bring new ideas into the subject. Although the asymptotic behavior
of such state is very well studied \cite{largex}, not so much is
known about the properties of the corresponding wavefunction
$\Psi(x)$ near the origin. It is expected that the explicit form of
the large $x$ and the small $x$ dependence of $\Psi(x)$ together
with some symmetry arguments should fix the entire state uniquely.
Following this argument one would like to determine the first few
terms of the Taylor expansion of $\Psi(x)$, the problem which was
addressed in Ref \cite{construction}. Since the wavefunction $\Psi(x)$ must be $Spin(9) \times SU(2)$ invariant \cite{hoppeinv}, 
the first term of the expansion $\phi := \Psi(0)$ is the $Spin(9) \times SU(2)$
singlet  depending only on the fermionic variables
$\theta_{\alpha A}$, $\{\theta_{\alpha A},\theta_{\beta B}\}=\delta_{AB}\delta_{\alpha\beta}$, ($\alpha=1,\ldots,16$, $A=1,2,3$). It turns out
that $\phi$ can be expressed in terms of elements of representations
of $SO(9)$, in a simple, closed form \cite{construction}. This
result agrees with earlier approach \cite{wosiek} where $\phi$ was
constructed using a different method. In both \cite{construction}
and \cite{wosiek}, the uniqueness of $\phi$ is argued relying on the
symbolic computer programme. In this paper we give a paper-pencil
prove of the uniqueness of $\phi$ using some novel intertwining
relations and identities involving $16 \times 16$ gamma matrices in
$9+1$ dimensions.

\section{The construction of $\phi$}

For fixed color index $A$, the sixteen fermions $\theta_{\alpha A}$, 
give rise to the 256 dimensional Hilbert space $H_{256}$,
correspondingly the total Hilbert space $H$ can be written as
$H=H_{256} \otimes H_{256} \otimes H_{256}$.  The $\textbf{256}$
representation of $SO(9)$ is reducible,
$\textbf{256}=\textbf{44}\oplus \textbf{84}\oplus\textbf{128}$, and
we find that among the possible tensor products, the relevant ones
(i.e. those involving the $SO(9)$ singlet) belong to
\cite{construction,slansky}
\begin{equation}
\textbf{44} \otimes \textbf{44} \otimes \textbf{44},  \label{ten1}
\end{equation}
\begin{equation}
\textbf{84} \otimes \textbf{84} \otimes \textbf{84},   \label{ten2}
\end{equation}
\begin{equation}
\textbf{44} \otimes \textbf{84} \otimes \textbf{84}, \ \textbf{84}
\otimes \textbf{84} \otimes \textbf{44}, \ \textbf{84} \otimes
\textbf{44} \otimes \textbf{84},  \label{ten3}
\end{equation}
\begin{equation}
\textbf{44} \otimes \textbf{128} \otimes \textbf{128}, \
\textbf{128} \otimes \textbf{44}  \otimes \textbf{128}, \
\textbf{128} \otimes \textbf{128} \otimes \textbf{44}\label{ten4}
\end{equation}
and
\begin{equation}
\textbf{84} \otimes \textbf{128} \otimes \textbf{128}, \
\textbf{128} \otimes \textbf{84} \otimes \textbf{128}, \
\textbf{128} \otimes \textbf{128} \otimes \textbf{84}. \label{ten5}
\end{equation}
There are in total fourteen $SO(9)$ singlets; 1, 1, 3, 3 and 6
corresponding to (\ref{ten1}), (\ref{ten2}), (\ref{ten3}), 
(\ref{ten4}) and (\ref{ten5}) respectively (note the  double
multiplicity in (\ref{ten5}) coming from the fact that $\textbf{128}
\otimes \textbf{128}$ gives two $\textbf{84}$'s while $\textbf{84}
\otimes \textbf{84}$ contains a singlet). An appropriate $SU(2)$
invariant combination of the 14 states yields the desired $Spin(9)
\times SU(2)$ singlet.

As shown in \cite{construction}, among the elements of five
representations in (\ref{ten1}), (\ref{ten2}) and (\ref{ten3}) there
exists only one such state, explicitly
\begin{equation}
\phi := \mathop{|||}_{44} 1 \rangle
+\frac{13}{36}\mathop{|||}_{844} 1 \rangle, \label{state}
\end{equation}
\[
\mathop{|||}_{44}1\rangle :=|su \rangle_1 |tu \rangle_2 |st \rangle_3,
\]
\[
\mathop{|||}_{844}1\rangle := |suv \rangle_1 |tuv \rangle_2 |st
\rangle_3 +   |tuv \rangle_1 |st \rangle_2 |suv \rangle_3  +   |st
\rangle_1 |suv \rangle_2 |tuv \rangle_3,
\]
where $|su \rangle_A$ and $|suv \rangle_A$ are the elements of the
$\textbf{44}$ and the $\textbf{84}$ representations, respectively (the elements of
(\ref{ten3}) do not contribute to $\phi$). The main result of this
paper is the proof of the fact that in the whole 14 dimensional
space of $SO(9)$ singlets, $\phi$ is a unique $Spin(9) \times SU(2)$
invariant state.

\section{The uniqueness}
Apart from the five $SO(9)$ singlets considered in \cite{construction}
\begin{equation}
S_1 := |su \rangle_1 |tu \rangle_2 |st \rangle_3,
\end{equation}
\begin{equation}
S_2 := \epsilon^{stupqrabc}|stu \rangle_1 |pqr \rangle_2 |abc
\rangle_3, 
\end{equation}
\[
S_3 := |suv \rangle_1 |tuv \rangle_2 |st\rangle_3, \ \ \ \ 
S_4 :=|tuv \rangle_1 |st \rangle_2 |suv \rangle_3, 
\]
\begin{equation}
S_5 := |st \rangle_1 |suv \rangle_2 |tuv \rangle_3
\end{equation}
corresponding to (\ref{ten1}), (\ref{ten2}) and (\ref{ten3})
respectively, there are 9 additional ones involving the
$\textbf{128}$ representation. Our choice is
\[
S_6:=|s\alpha \rangle_1 |t\alpha \rangle_2 |st \rangle_3, \ \ \ \
S_7:=|s\alpha \rangle_1 |st \rangle_2 |t\alpha \rangle_3,
\]
\begin{equation}
S_8:=|st \rangle_1 |s\alpha \rangle_2 |t\alpha \rangle_3, \label{S1}
\end{equation}
\[
S_9:=\gamma^s_{\alpha\beta}|u\alpha \rangle_1 |v\beta \rangle_2 |suv \rangle_3, \ \ \ \
S_{10}:=\gamma^s_{\alpha\beta}|u\alpha \rangle_1 |suv \rangle_2 |v\beta \rangle_3,
\]
\begin{equation}
S_{11}:=\gamma^s_{\alpha\beta}|suv \rangle_1 |u\alpha \rangle_2 |v\beta \rangle_3,  \label{S2}
\end{equation}
\[
S_{12}:=\gamma^{suv}_{\alpha\beta}|t\alpha \rangle_1 |t\beta \rangle_2 |suv \rangle_3, \ \ \ \
S_{13}:=\gamma^{suv}_{\alpha\beta}|t\alpha \rangle_1 |suv \rangle_2 |t\beta \rangle_3,
\]
\begin{equation}
S_{14}:=\gamma^{suv}_{\alpha\beta}|suv \rangle_1 |t\alpha \rangle_2 |t\beta \rangle_3,  \label{S3}
\end{equation}
(although the states in (\ref{S2}) and (\ref{S3}) have the same
representation content, they are linearly independent since
there does not exist an identity such as $\delta^{rt}\gamma^{suv}_{\alpha\beta}
\propto \delta^{tu}\delta^{rv}\gamma^s_{\alpha\beta} $  ).

The $SU(2)$ invariance of the linear combination $\tilde{\phi}:=
\sum_i a_i S_i$, $a_i \in \mathbb{C}$, implies that $J_A\tilde{\phi}
= 0 $, $A=1,2,3$ where $J_A$ are the $SU(2)$ generators
$J_A=\frac{1}{2}\epsilon_{ABC}\theta_{\alpha B}\theta_{\alpha C}$.
Let us take  $A=3$ and denote the matrix representation of $J_3$ by
$J_{ij}$, i.e. $J_3 S_i =\sum_j J_{ji}S_j$. The $SU(2)$ invariance
is now equivalent to the matrix equation
\begin{equation}
\sum_i a_i J_{ji} = 0,
\end{equation}
hence a uniqueness of $\phi$ is equivalent to the existence of a
unique eigenvector of matrix $J$ corresponding to the zero
eigenvalue.

\noindent The first 5 rows of the matrix $J_{ij}$ can be determined from  \cite{construction}
\[
J_3S_1 = \frac{13}{4}S_6, \ \ \ \  J_3S_2 = -\frac{3456}{5}S_9+ \frac{972}{5}S_{12}, \ \ \ \ J_3S_3 = -9S_6,
\]
\[
J_3S_4 = \frac{13 i}{\sqrt{2}}S_9  -\frac{23 i}{\sqrt{2}}S_{12}, \ \
\ \ J_3S_5 =-\frac{13 i}{\sqrt{2}}S_9 + \frac{23 i}{\sqrt{2}}S_{12}
\ \ .
\]
In deriving the above result it is essential to take advantage of
the intertwining relations
\begin{equation}
2\theta_{\alpha A}|st \rangle_A = \gamma^s_{\alpha\beta}|t \beta \rangle_A +  \gamma^t_{\alpha\beta}|s \beta \rangle_A , \label{int1}
\end{equation}
\begin{equation}
\theta_{\alpha A}|stu \rangle_A  = \frac{i}{\sqrt{2}}\left(
\gamma^{st}_{\alpha \beta}|u\beta \rangle_A + \gamma^{us}_{\alpha
\beta}|t\beta \rangle_A +\gamma^{tu}_{\alpha \beta}|s\beta \rangle_A
\right), \label{int2}
\end{equation}
note however that, because of the appearance of the $\textbf{128}$
in (\ref{S1}), (\ref{S2}) and (\ref{S3}), the evaluation of $J_{ij}$
for $j>5$ requires one more relation, namely \footnote{Eqn.
(\ref{int3}) was also known to J.Hoppe and D. Lundholm.}
\begin{equation}
\theta_{\alpha,A}|t\beta \rangle_A =
\frac{1}{2}\gamma^{\mu}_{\alpha\beta}|ut\rangle_A -
\frac{i}{36\sqrt{2}}\gamma^{tsuv}_{\alpha\beta}|suv \rangle_A -
\frac{i}{6\sqrt{2}}\gamma^{uv}_{\alpha\beta}|uvt\rangle_A
\label{int3}.
\end{equation}
After some algebra (see the next section for the details) we find that
\[
J_3S_6 = 4 S_1 + \frac{5}{9} S_3, \ \ \ \  J_3S_7 = \frac{1}{2}S_8-\frac{4i}{3\sqrt{2}}S_{11}-\frac{5i}{36 \sqrt{2}}S_{14},
\]
\[
J_3S_8 =\frac{1}{2}S_7-\frac{4i}{3\sqrt{2}}S_{10}-\frac{5i}{36 \sqrt{2}}S_{13}, \ \ \ \
 J_3S_9 = \frac{1}{162}S_2-\frac{8i}{3\sqrt{2}}S_4-\frac{8i}{3 \sqrt{2}}S_5,
\]
\[
J_3S_{10} = \frac{7}{6}S_{11}-\frac{1}{18}S_{14}, \ \ \ \ J_3S_{11} = \frac{7}{6}S_{10}-\frac{1}{18}S_{13} , \ \ \ \
\]
\[
J_3S_{12}=-\frac{11}{162}S_{2}+\frac{8i}{3\sqrt{2}}S_4+\frac{8i}{3\sqrt{2}}S_5,
\]
\[
J_3S_{13}=-\frac{126i}{\sqrt{2}}S_8-62S_{11}, \ \ \ \ J_3S_{14}=-\frac{126i}{\sqrt{2}}S_7-62S_{10},
\]
hence
 \[
 [J]_{ij}=\left[ \begin{smallmatrix}
 0 & 0 & 0 & 0 & 0 & 4 & 0 & 0 & 0 & 0 & 0 & 0 & 0 & 0 \\
 0 & 0 & 0 & 0 & 0 & 0 & 0 & 0 & \frac{1}{162} & 0 & 0 & -\frac{11}{162} & 0 & 0 \\
 0 & 0 & 0 & 0 & 0 & \frac{5}{9} & 0 & 0 & 0 & 0 & 0 & 0 & 0 & 0 \\
 0 & 0 & 0 & 0 & 0 & 0 & 0 & 0 & -\frac{8 i }{3\sqrt{2}} & 0 & 0 & \frac{8 i} {3\sqrt{2}} & 0 & 0 \\
 0 & 0 & 0 & 0 & 0 & 0 & 0 & 0 & -\frac{8 i}{3\sqrt{2}} & 0 & 0 & \frac{8 i} {3\sqrt{2}} & 0 & 0 \\
 \frac{13}{4} & 0 & -9 & 0 & 0 & 0 & 0 & 0 & 0 & 0 & 0 & 0 & 0 & 0 \\
 0 & 0 & 0 & 0 & 0 & 0 & 0 & \frac{1}{2} & 0 & 0 & 0 & 0 & 0 & -63 i \sqrt{2} \\
 0 & 0 & 0 & 0 & 0 & 0 & \frac{1}{2} & 0 & 0 & 0 & 0 & 0 & -63 i \sqrt{2} & 0 \\
 0 & -\frac{3456}{5} & 0 & \frac{13 i}{\sqrt{2}} & -\frac{13 i}{\sqrt{2}} & 0 & 0 & 0 & 0 & 0 & 0 & 0 & 0 & 0 \\
 0 & 0 & 0 & 0 & 0 & 0 & 0 & -\frac{4i}{3\sqrt{2}} & 0 & 0 & \frac{7}{6} & 0 & 0 & -62 \\
 0 & 0 & 0 & 0 & 0 & 0 & -\frac{4i}{3\sqrt{2}} & 0 & 0 & \frac{7}{6} & 0 & 0 & -62 & 0 \\
 0 & \frac{972}{5} & 0 & -\frac{23 i}{\sqrt{2}} & \frac{23 i}{\sqrt{2}} & 0 & 0 & 0 & 0 & 0 & 0 & 0 & 0 & 0 \\ 
 0 & 0 & 0 & 0 & 0 & 0 & 0 & -\frac{5 i}{36 \sqrt{2}} & 0 & 0 & -\frac{1}{18} & 0 & 0 & 0 \\
 0 & 0 & 0 & 0 & 0 & 0 & -\frac{5 i}{36 \sqrt{2}} & 0 & 0 & -\frac{1}{18} & 0 & 0 & 0 & 0
\end{smallmatrix} \right].
\]
The matrix $J$ can be easily diagonalized and we find that its
kernel is two dimensional spanned by vectors $S_1+\frac{13}{36}S_3$
and $S_4+S_5$. Since the singlet must be invariant with respect to
any permutation of the color index $A$, the only possibility is the
cyclically invariant combination given by (\ref{state}).

\subsection{Detailed calculation}

Below we present the evaluation of $J_3=\theta_{\alpha
1}\theta_{\alpha 2}$ acting on $S_i$, $i>5$ focusing on most
important parts of the calculation.

For $i=6$ the state $J_3S_6=\theta_{\alpha 1}\theta_{\alpha
2}|s\beta \rangle |t\beta \rangle |st\rangle $ consists of 9 terms
(c.p. (\ref{int3})), explicitly
\[
\frac{1}{4}Tr(\gamma^{u}\gamma^{u_1}) |us \rangle |u_1 t \rangle |st
\rangle = 4 S_1, \ \ \ \ -\frac{i}{72\sqrt{2}}Tr(\gamma^{u}
\gamma^{t p q r}) |us \rangle |pqr \rangle |st \rangle = 0,
\]
\[
-\frac{i}{12\sqrt{2}}Tr(\gamma^{u} \gamma^{p q}) |us \rangle |pqt
\rangle |st \rangle = 0, \ \ \ \
-\frac{i}{72\sqrt{2}}Tr(\gamma^{spqr} \gamma^{u}) |pqr \rangle |ut
\rangle |st \rangle = 0,
\]
\[
-\frac{1}{2592}Tr(\gamma^{spqr} \gamma^{tp_1q_1r_1}) |pqr \rangle |p_1q_1r_1 \rangle |st \rangle = \frac{1}{9}S_3,
\]
\[
\frac{1}{432}Tr(\gamma^{spqr} \gamma^{p_1q_1})  |pqr \rangle |p_1q_1t \rangle |st \rangle = 0,
\]
\[
-\frac{i}{12\sqrt{2}}Tr(\gamma^{pq} \gamma^{u}) |pqs \rangle |ut \rangle |st \rangle = 0,
\]
\[
-\frac{1}{432}Tr(\gamma^{pq} \gamma^{tp_1q_1r_1}) |pqs \rangle |p_1q_1r_1 \rangle |st \rangle = 0,
\]
\[
-\frac{1}{72}Tr(\gamma^{pq} \gamma^{p_1 q_1}) |pqs \rangle |p_1q_1t \rangle |st \rangle = \frac{4}{9}S_3,
\]
which were evaluated using the following identities
\[
Tr(\gamma^{u} \gamma^{t p q r})=0, \ \ \ \ Tr(\gamma^{u} \gamma^{p q})=0, \ \ \ \
Tr(\gamma^{spqr} \gamma^{u})=0, \ \ \ \ Tr(\gamma^{spqr} \gamma^{p_1q_1})=0,
\]
\[
Tr(\gamma^{spqr} \gamma^{tp_1q_1r_1})=16 \sum_{\pi \in S_4}{sgn(\pi)\delta^{s \pi(t)}\delta^{p \pi (p_1)}\delta^{q \pi(q_1)}\delta^{r \pi(r_1)}},
\]
\[
Tr(\gamma^{pq} \gamma^{u})=0, \ \ \ \ Tr(\gamma^{pq} \gamma^{tp_1q_1r_1})=0, \ \ \ \
Tr(\gamma^{pq} \gamma^{p_1 q_1})=-16(\delta^{pp_1}\delta^{qq_1}-\delta^{pq_1}\delta^{qp_1}).
\]
Therefore we have
\begin{equation}
J_3 S_6 = 4S_1+\frac{5}{9}S_3  \label{js6}.
\end{equation}

For $i=7$ the state $J_3S_7=\theta_{\alpha 1}\theta_{\alpha 2}|s\beta \rangle |st \rangle |t\beta \rangle $ consists of 6 terms. They are
\[
\frac{1}{4}[\gamma^u\gamma^s]_{\beta\beta^{'}} |us\rangle |t\beta^{'} \rangle |t\beta \rangle = 0, \ \ \ \
\frac{1}{4}[\gamma^u\gamma^t]_{\beta\beta^{'}}|us\rangle |s\beta^{'} \rangle |t\beta \rangle = \frac{1}{2}S_8
\]
\[
\frac{i}{72\sqrt{2}}[\gamma^{spqr}\gamma^s]_{\beta\beta^{'}}|pqr\rangle |t\beta^{'} \rangle |t\beta \rangle = -\frac{i }{12\sqrt{2}}S_{14},
\]
\[
\frac{i}{12\sqrt{2}}[\gamma^{pq}\gamma^s]_{\beta\beta^{'}}|pqs\rangle |t\beta^{'} \rangle |t\beta \rangle =  -\frac{i}{12\sqrt{2}}S_{14},
\]
\[
-\frac{i}{72\sqrt{2}}[\gamma^{spqr}\gamma^t]_{\beta\beta^{'}}|pqr\rangle |s\beta^{'} \rangle |t\beta \rangle = -\frac{i}{\sqrt{2}}S_{11}+\frac{i}{36 \sqrt{2}}S_{14},
\]
\[
-\frac{i}{12\sqrt{2}}[\gamma^{pq}\gamma^t]_{\beta\beta^{'}}|pqs\rangle |s\beta^{'} \rangle |t\beta \rangle = -\frac{i}{3\sqrt{2}}S_{11}
\]
where we used the identities
\[
\gamma^{spqr}\gamma^s = -6 \gamma^{pqr}, \ \ \ \ [\gamma^{spqr},\gamma^t]=2(\delta^{tr}\gamma^{spq}-\delta^{tq}\gamma^{spr}+\delta^{tp}\gamma^{sqr}-\delta^{ts}\gamma^{pqr})
\]
\[
[\gamma^{pq},\gamma^t]=2\gamma^p\delta^{qt}-2\gamma^q\delta^{pt}, \ \ \ \ \gamma^u\gamma^s=\delta^{us}\textbf{1}+\gamma^{us}
\]
the constraint $\sum_s| ss \rangle =0$ and the Rarita-Schwinger
constraint  $\gamma^s_{\alpha\beta}|t\beta \rangle=0$. Therefore we
obtain
\begin{equation}
J_3 S_7 = \frac{1}{2}S_8-\frac{4i}{3\sqrt{2}}S_{11}-\frac{5i}{36 \sqrt{2}}S_{14} \label{js7}
\end{equation}
The evaluation  of $J_3 S_8$ is analogous to $J_3 S_7$ and we find that
\begin{equation}
J_3 S_8 = \frac{1}{2}S_7-\frac{4i}{3\sqrt{2}}S_{10}-\frac{5i}{36 \sqrt{2}}S_{13}  \label{js8}
\end{equation}

For $i=9$ the state $J_3S_9=\theta_{\alpha' 1}\theta_{\alpha' 2}\gamma^{s}_{\alpha \beta}|u\alpha \rangle |v\beta \rangle |suv\rangle $ consists of 9 terms. They are
\[
\frac{1}{4}Tr(\gamma^p\gamma^q\gamma^r)|ps \rangle|rt \rangle|qst \rangle=0, \ \ \ \
-\frac{i}{72\sqrt{2}}Tr(\gamma^{pqrs}\gamma^t\gamma^u)|qrs \rangle|uv \rangle|tpv \rangle=0,
\]
\[
-\frac{i}{12\sqrt{2}}Tr(\gamma^{pq}\gamma^r\gamma^s)|pqt\rangle|su\rangle |rtu\rangle=-\frac{8i}{3 \sqrt{2} }S_4,
\]
\[
-\frac{i}{72\sqrt{2}}Tr(\gamma^p\gamma^q\gamma^{rstu})|pv\rangle|stu\rangle|qvr\rangle=0,
\]
\[
\frac{1}{2592} Tr(\gamma^{pqrs}\gamma^{t}\gamma^{p_1q_1r_1s_1})|qrs\rangle |q_1r_1s_1\rangle |tpp_1\rangle=\frac{1}{162}S_2,
\]
\[
\frac{1}{432}Tr(\gamma^{pp_1}\gamma^q\gamma^{rstu})|pp_1v\rangle|stu\rangle|qvr\rangle=0,
\]
\[
-\frac{i}{12\sqrt{2}} Tr(\gamma^r\gamma^s\gamma^{pq})|ru\rangle|pqt\rangle |uts\rangle=-\frac{8i}{3 \sqrt{2} }S_5,
\]
\[
\frac{1}{432 } Tr(\gamma^{pqrs}\gamma^t\gamma^{p_1q_1})|qrs\rangle|p_1q_1v\rangle|tpv\rangle,
\]
\[
\frac{1}{72 } Tr(\gamma^{pq}\gamma^r\gamma^{st})|pqu\rangle|stv\rangle|ruv\rangle=0,
\]
where we used the identities
\[
Tr(\gamma^p\gamma^q\gamma^r) = 0, \ \ \ \ Tr(\gamma^{pqrs}\gamma^t\gamma^u)= 0,
\]
\[
Tr(\gamma^{pq}\gamma^r\gamma^s)=Tr(\gamma^p\gamma^q\gamma^{rs})=-16(\delta^{rp}\delta^{sq}-\delta^{rq}\delta^{sp}),
\]
\[
Tr(\gamma^p\gamma^q\gamma^{rstu})=0, \ \ \ \
Tr(\gamma^{pqrs}\gamma^{t}\gamma^{p_1q_1r_1s_1})=16\epsilon^{pqrstp_1q_1r_1s_1},
\]
\[
Tr(\gamma^{pp_1}\gamma^q\gamma^{rstu})=0, \ \ \ \
Tr(\gamma^{pqrs}\gamma^t\gamma^{p_1q_1})=0, \ \ \ \
Tr(\gamma^{pq}\gamma^r\gamma^{st})=0.
\]
Therefore we obtain
\begin{equation}
J_3S_9 = \frac{1}{162}S_2-\frac{8i}{3\sqrt{2}}S_4-\frac{8i}{3 \sqrt{2}}S_5 \label{js9}.
\end{equation}

For $i=10$ the state $J_3S_{10}=\theta_{\alpha' 1}\theta_{\alpha' 2}\gamma^{s}_{\alpha \beta}|u\alpha \rangle|suv\rangle |v\beta \rangle  $ consists of 9 terms.
We use the Rarita-Schwinger constraint and
\[\gamma^{ts}\gamma^s=8\gamma^t, \ \ \ \  \gamma^u\gamma^s=2\delta^{us}\textbf{1}+\gamma^{us}, \ \ \ \ \gamma^{tabc}=\gamma^t\gamma^{abc}, t\neq{a,b,c}
\]
to find that they are
\[
\frac{i}{2 \sqrt{2}}[\gamma^{pq}\gamma^r\gamma^s]_{\alpha\beta}|rq\rangle|t\alpha\rangle|t\beta\rangle=0, \ \ \ \
\frac{1}{72}[\gamma^{pq}\gamma^{qstu}\gamma^p]_{\alpha\beta}|stu\rangle |v\alpha\rangle|v\beta\rangle=-\frac{1}{9}S_{14},
\]
\[
\frac{1}{12}[\gamma^{pq}\gamma^{rs}\gamma^p]_{\alpha\beta}|rsq\rangle |v\alpha\rangle|v\beta\rangle=-\frac{1}{3}S_{14}, \ \ \ \
\frac{i}{2 \sqrt{2}}[\gamma^{pq}\gamma^{r}\gamma^q]_{\alpha\beta}|rs\rangle |s\alpha\rangle|p\beta\rangle=0,
\]
\[
\frac{1}{72}[\gamma^{pq}\gamma^{rstu}\gamma^q]_{\alpha\beta}|stu\rangle |r\alpha\rangle|p\beta\rangle=0, \ \ \ \
\frac{1}{12}[\gamma^{pq}\gamma^{rs}\gamma^q]_{\alpha\beta}|rst\rangle |t\alpha\rangle|p\beta\rangle=2S_{11},
\]
\[
\frac{i}{2 \sqrt{2}}[\gamma^{pq}\gamma^{r}\gamma^s]_{\alpha\beta}|rp\rangle |s\alpha\rangle|q\beta\rangle=0, \ \ \ \
\frac{1}{72}[\gamma^{pq}\gamma^{prst}\gamma^u]_{\alpha\beta}|rst\rangle |u\alpha\rangle|q\beta\rangle=\frac{2}{9}S_{14},
\]
\[
\frac{1}{12}[\gamma^{pq}\gamma^{rs}\gamma^t]_{\alpha\beta}|rsp\rangle |t\alpha\rangle|q\beta\rangle=\frac{1}{6}S_{14}-\frac{5}{6}S_{11}.
\]
Therefore we obtain
\begin{equation}
J_3S_{10} = \frac{7}{6}S_{11}-\frac{1}{18}S_{14} \label{js10}.
\end{equation}
Evaluation of $J_3S_{11}$ is analogous to $J_3S_{10}$ and we find that
\begin{equation}
J_3S_{11} = \frac{7}{6}S_{10}-\frac{1}{18}S_{13} \label{js11}.
\end{equation}

For $i =12$ the state $J_3S_{12}=\theta_{\alpha' 1}\theta_{\alpha'
2}\gamma^{suv}_{\alpha \beta}|t\alpha \rangle|t\beta\rangle |suv\rangle  $ consists of 9 terms.
They are
\[
\frac{1}{4}Tr(\gamma^p\gamma^{qrs}\gamma^t)|pa\rangle|ta \rangle|qrs\rangle=0,
\]
\[
-\frac{i}{72 \sqrt{2}}Tr(\gamma^{pqrs}\gamma^{p_1q_1r_1}\gamma^u)|qrs \rangle|up \rangle|p_1q_1r_1\rangle=-\frac{16i }{3\sqrt{2}}S_4,
\]
\[
-\frac{i}{12 \sqrt{2}}Tr(\gamma^{pq}\gamma^{qst}\gamma^u)|pqt\rangle|tu\rangle|qst\rangle=\frac{8 i} {\sqrt{2}}S_4,
\]
\[
-\frac{i}{72 \sqrt{2}}Tr(\gamma^{p}\gamma^{qrs}\gamma^{tq_1r_1s_1})|pt\rangle|q_1r_1s_1\rangle|qrs\rangle=-\frac{16 i }{3\sqrt{2}}S_5,
\]
\[
-\frac{1}{2592}Tr(\gamma^{pqrs}\gamma^{tuv}\gamma^{pq_1r_1s_1})|qrs\rangle |q_1r_1s_1\rangle |tuv\rangle=\frac{1}{162}S_2,
\]
\[
-\frac{1}{432}Tr(\gamma^{pq}\gamma^{rst}\gamma^{ur_1s_1t_1})|pqu\rangle|r_1s_1t_1\rangle|rst\rangle=-\frac{1}{27}S_2,
\]
\[
-\frac{i}{12 \sqrt{2}}Tr(\gamma^p\gamma^{qrs}\gamma^{tu})|pv\rangle|tuv\rangle|qrs\rangle=\frac{8 i}{\sqrt{2}}S_5,
\]
\[
-\frac{1}{432}Tr(\gamma^{pqrs}\gamma^{q_1r_1s_1}\gamma^{tu})|qrs\rangle|tup\rangle|q_1r_1s_1\rangle=-\frac{1}{27}S_2,
\]
\[
-\frac{1}{72}Tr(\gamma^{pq}\gamma^{rst}\gamma^{uv})|pqw\rangle|uvw\rangle|rst\rangle=0,
\]
where we used
\[
Tr(\gamma^p\gamma^{qrs}\gamma^t)=0, \ \ \ \
\]
\[
Tr(\gamma^{pqrs}\gamma^{p_1q_1r_1}\gamma^u)=Tr(\gamma^u\gamma^{p_1q_1r_1}\gamma^{pqrs})=
16 \sum_{\pi \in S_4}sgn(\pi)\delta^{s\pi(u)u}\delta^{p\pi(p_1)}\delta^{q\pi(q_1)}\delta^{r\pi(r_1)},
\]
\[
Tr(\gamma^{pqrs}\gamma^{tuv}\gamma^{pq_1r_1s_1})=-48\epsilon^{qrstuvq_1r_1s_1}, \ \ \ \
Tr(\gamma^{pq}\gamma^{rst}\gamma^{ur_1s_1t_1})=16\epsilon^{pqur_1s_1t_1rst},
\]
\[ Tr(\gamma^{rst}\gamma^u\gamma^{pq})=Tr(\gamma^{rst}\gamma^{pq}\gamma^u)=-16\sum_{\pi \in S_3}{\delta^{p\pi(r)}\delta^{q\pi(s)}\delta^{u\pi(t)}},
\]
\[
Tr(\gamma^{pqrs}\gamma^{q_1r_1s_1}\gamma^{tu})=16\epsilon^{qrstupq_1r_1s_1}, \ \ \ \ Tr(\gamma^{pq}\gamma^{rst}\gamma^{uv})=0.
\]
Therefore we obtain
\begin{equation}
J_3S_{12}=-\frac{11}{162}S_{2}+\frac{8i}{3\sqrt{2}}S_4+\frac{8i}{3\sqrt{2}}S_5 \label{js12}.
\end{equation}

For $i=13$ state $J_3S_9=\theta_{\alpha'1}\theta_{\alpha'2}\gamma^{suv}_{\alpha \beta}|t\alpha \rangle|suv\rangle |t\beta\rangle  $ consists of 3 terms.
They are
\[
\frac{3i}{2 \sqrt{2}}[\gamma^{pq}\gamma^{r}\gamma^{pqu}]_{\alpha\beta}|rt \rangle|u\alpha\rangle |t\beta\rangle=-\frac{126i}{\sqrt{2}}S_8,
\]
\[
\frac{3}{72}[\gamma^{pq}\gamma^{rstu}\gamma^{pqv}]_{\alpha\beta}|stu \rangle|v\alpha\rangle |r\beta\rangle=0,
\]
\[
\frac{3}{12 }[\gamma^{pq}\gamma^{rs}\gamma^{pqv}]_{\alpha\beta}|rst \rangle|v\alpha\rangle |t\beta\rangle=-62S_{11},
\]
where we used the R-S constraint and
\[
[\gamma^{su},\gamma^{v}]=2\gamma^{s}\delta^{vt}-2\gamma^{u}\delta^{ut}, \ \ \ \ \gamma^{su}\gamma^{suv}=-56\gamma^{v},
\]
\[
\gamma^{tabc}=\gamma^t\gamma^{abc}, t\neq{a,b,c}, \ \ \ \ \gamma^{abct}=\gamma^{abc}\gamma^t, t\neq{a,b,c}, \ \ \ \ [\gamma^{t},\gamma^{suv}]=2\gamma^{tsuv},
\]
\[
[\gamma^{su},\gamma^{ab}]=2(\gamma^{sb}\delta^{ua}-\gamma^{sa}\delta^{ub}+\gamma^{ua}\delta^{sb}-\gamma^{ub}\delta^{sa}).
\]
Therefore we obtain
\begin{equation}
J_3S_{13}=-\frac{126i}{\sqrt{2}}S_8-62S_{11}\label{js13}.
\end{equation}
Evaluation of $J_3S_{14}$ is analogous to $J_3S_{14}$
\begin{equation}
J_3S_{14}=-\frac{126i}{\sqrt{2}}S_7-62S_{10}\label{js14}.
\end{equation}

\section{Outlook}

The uniqueness of the $Spin(9)\times SU(2)$ state $\phi$, proven in
this paper,  is a starting point for the unique Fock space
representation of the hamiltonian of the full matrix model
\[
H= K+V+H_F
\]
\[
K=-\partial_{As}\partial_{As}, \ \ \ \ V=\frac{1}{2}(\epsilon_{ABC}x_{Bs}x_{Ct})^2, \ \ \ \ H_F= if_{CAB}\gamma^s_{\alpha\beta}x_{Cs}\theta_{A \alpha}\theta_{B \beta},
\]
in terms of $Spin(9)\times SU(2)$ invariant basis. To be more specific consider the normalized "vacuum" state
\[
|v \rangle := \frac{1}{\left\| \phi \right\|}|0\rangle_B \otimes \phi,
\]
where $\left\| \phi \right\|^2=14014/9$ (explicitly checked on the
computer)  and $| 0 \rangle_B$ is the bosonic Fock vacuum (in the
coordinate representation $\langle x|0 \rangle \propto exp(-\frac{1}{2}x_{As}x_{As})$).
Such state is also $Spin(9)\times SU(2)$ invariant and gives a
possibility to represent $H$ in the $Spin(9)\times SU(2)$ invariant
basis obtained by acting with bosonic creation operators
$a^{\dagger}_{As}$ and fermionic operators $\theta_{A \alpha}$ on
$|v \rangle$. 

First step towards this direction can be done by finding the
expectation  value $\langle v |H| v \rangle$. There is no
contribution from $H_F$ since $H_F$ is linear in $x_{As}$ while the
contributions from $K$ and $V$ are $27/2$ and $54$ respectively,
implying that
\[
\langle v |H| v \rangle = 67.5,
\]
a rather large number considering the existence of the (conjectured)
zero-energy ground state of the model.

\section*{Acknowledgments}
We thank J. Hoppe and D. Lundholm for discussions.


\begin{thebibliography}{7}

\bibitem{model}
W. Taylor, \emph{M(atrix) Theory: Matrix Quantum Mechanics as a Fundamental Theory}, Rev. Mod. Phys. 73 (2001) 419-462,  {\tt arXiv:hep-th/0101126},
\\
J. Hoppe, \textit{Membranes and Matrix Models}, {\tt arXiv:hep-th/0206192}.

\bibitem{largex}

J. Fr\"ohlich, J. Hoppe, \textit{On Zero-Mass Ground States in Super-Membrane Matrix Models},
{\tt arXiv:hep-th/9701119};

J. Hoppe, \textit{On the Construction of Zero Energy States in Supersymmetric Matrix Models}, {\tt arXiv:hep-th/9709132};

M. B. Halpern, C. Schwartz, \textit{Asymptotic Search for Ground States of SU(2) Matrix Theory}, Int. J. Mod. Phys. A13 (1998) 4367-4408, {\tt arXiv:hep-th/9712133};

A. Konechny, \textit{On Asymptotic Hamiltonian for SU(N) Matrix Theory}, JHEP 9810 (1998) 018, {\tt arXiv:hep-th/9805046};

M. Bordemann, J. Hoppe, R. Suter, \textit{Zero Energy States for SU(N): A Simple Exercise in Group Theory ?}, {\tt arXiv:hep-th/9909191};

J. Hoppe, \textit{Asymptotic Zero Energy States for SU(N greater or equal 3)},
{\tt arXiv:hep-th/9912163};

J. Fr\"ohlich, G. M. Graf, D. Hasler, J. Hoppe, S.-T. Yau, \textit{Asymptotic form of zero energy wave functions in supersymmetric matrix models,} Nucl. Phys. B567 (2000) 231-248, {\tt arXiv:hep-th/9904182};

J. Hoppe, J. Plefka, \textit{The Asymptotic Groundstate of SU(3) Matrix Theory},
{\tt arXiv:hep-th/0002107};

D. Hasler, J. Hoppe, \textit{Asymptotic Factorisation of the Ground-State for SU(N)-invariant Supersymmetric Matrix-Models}, {\tt arXiv:hep-th/0206043};

V. Bach, J. Hoppe, D. Lundholm, \textit{Dynamical Symmetries in Supersymmetric Matrix Models}, {\tt arXiv:hep-th/07060355};

J. Hoppe, D. Lundholm, \textit{On the Construction of Zero Energy States in Supersymmetric Matrix Models IV}, {\tt arXiv:0706.0353};

\bibitem{construction} J. Hoppe, D. Lundholm, M. Trzetrzelewski, \textit{Construction of the Zero-Energy State of SU(2)-Matrix Theory: Near the Origin}, Nucl. Phys. B817:155-166, 2009 {\tt arXiv:0809.5270},

\bibitem{hoppeinv} D. Hasler, J. Hoppe, \textit{Zero Energy States of Reduced Super Yang-Mills Theories
in $d+1 = 4,6$ and 10 dimensions are necessarily $Spin(d)$ invariant,} {\tt arXiv:hep-th/0211226}.

\bibitem{wosiek} J. Wosiek, \textit{On the SO(9) structure of supersymmetric Yang-Mills quantum mechanics} Phys. Lett. B619 (2005) 171-176, {\tt arXiv:hep-th/0503236}.

\bibitem{slansky}  R. Slansky, \textit{Group theory for unified model building}, Phys. Rep., Vol. 79, Issue 1, p. 1-128 (1981).

\end{thebibliography}
\end{document}